\documentclass{elsart}

\newcommand{\trg}{{\rm trg}}
\newcommand{\Kscr}{{\mathcal K}}

\begin{document}

\begin{frontmatter}

\title{Does localization occur in a hierarchical random--matrix model
  for many--body states?\thanksref{DEDICATE}}

\author[HD]{Thomas Rupp\thanksref{CONTACT}}, 
\author[HD]{Hans A. Weidenm{\"u}ller}, and
\author[SB]{Jean Richert} 
\address[HD]{Max--Planck--Institut f{\"u}r
  Kernphysik, Postfach 103980, 69029 Heidelberg, Germany}
\address[SB]{Laboratoire de Physique Th\'eorique\thanksref{UMR},
  Universit\'e Louis Pasteur,\\ 3 rue de l'Universit\'e, 67084
  Strasbourg Cedex, France}

\thanks[DEDICATE]{dedicated to Franz Wegner on the occasion of his
  60th birthday}
\thanks[CONTACT]{electronic mail: thomas.rupp@mpi-hd.mpg.de}
\thanks[UMR]{Unit\'e Mixte de
  Recherche CNRS -- Universit\'e (UMR 7085)}

\begin{abstract}
  We use random--matrix theory and supersymmetry techniques to work
  out the two--point correlation function between states in a
  hierarchical model which employs Feshbach's chaining hypothesis:
  Classes of many--body states are introduced. Only states within the
  same or neighboring classes are coupled. We assume that the density
  of states per class grows monotonically with class index. The
  problem is mapped onto a one--dimensional non--linear sigma model.
  In the limit of a large number of states in each class we derive the
  critical exponent for the growth of the level density with class
  index for which delocalization sets in. From a realistic modelling
  of the class--dependence of the level density, we conclude that the
  model does not predict Fock--space localization in nuclei.
\end{abstract}


\end{frontmatter}

Statistical methods play an important role in the quantal many--body
problem. We recall the importance of the Gaussian Orthogonal Ensemble
of random matrices (the GOE) for the description of fluctuation
properties of the energies and wave functions of neutron resonances
and other data on nuclear energy levels located a few MeV above the
ground state. The use of the GOE is not restricted to nuclei, of
course, and this ensemble is equally important in other many--body
systems like atoms or quantum dots~\cite{guhr}.

It is well known that the GOE does not furnish a completely adequate
description of stochasticity in any of these systems. This is because
the forces between the constituents (nucleons in the case of nuclei,
electrons in the case of atoms and quantum dots) are predominantly of
two--body type. Then it is more natural to assume that the matrix
elements of the two--body interaction are random variables. Both in
nuclear and in atomic physics, there is empirical evidence that this
assumption is viable~\cite{gervois,flambaum}. The random--matrix
ensemble which implements this assumption in large shell--model spaces
(the two--body random ensemble) differs from the GOE: The number of
independent two--body matrix elements is generically much smaller than
the dimension of a typical shell--model matrix, while for the GOE, the
number of independent matrix elements is proportional to the square of
the dimension of the matrix. Numerical studies have indicated long ago
that this characteristic difference is immaterial for spectral
fluctuation properties~\cite{french,bohigas}. However, these studies
were limited to small matrix dimensions. Therefore, the question
persists whether the GOE and the random two--body ensemble are fully
equivalent.

The question has resurfaced with the recent observation that in
quantum dots, a statistical model for the two--body interaction
allows, in principle, for localization in Fock space~\cite{agkl}.
Numerical studies seem to support this proposal, at least at
sufficiently high energies~\cite{leyronas}, while near the Fermi
energy no evidence for localization was found~\cite{mejia}. It is then
only natural to ask whether a random two--body interaction in nuclei
might likewise lead to Fock--space localization in nuclei. Such
localization would have exciting implications for level statistics and
wave function properties. Moreover, Fock--space localization could
lead to novel aspects of nuclear reaction dynamics.

Localization is a wave phenomenon discovered by
Anderson~\cite{anderson}. It occurs when non--interacting electrons
move diffusively under the influence of impurity scattering through a
conductor. The Anderson tight--binding model describes this process in
terms of a lattice with random on--site energies distributed uniformly
over an interval of length $2 W$ and with fixed hopping matrix
elements $V$ connecting neighboring sites. Localized eigenfunctions of
the tight--binding Hamiltonian are not spread uniformly over the
entire lattice but possess an envelope which decays exponentially at
large distance. Anderson localization occurs not only in the passage
of Schr\"odinger waves through a disordered medium but likewise in the
passage of light through a medium with a disordered index of
refraction~\cite{wiersma}.

The concept of localization can be extended to many--body systems.
This~was first suggested in quantum chemistry~\cite{logan}. Logan and
Wolynes observed a close topological similarity between a statistical
description of the many--body problem in polyatomic molecules and the
tight--binding model. An analogous similarity exists in quantum
dots~\cite{agkl}. In the present paper, we point out that a
corresponding similarity also exists in a Fock--space description of
\mbox{many--body} systems like nuclei or atoms with random two--body
interactions. We investigate the question whether in these systems,
localization is expected to occur.

The occurrence of localization can either be established by very
extensive numerical calculations or analytically. Here, we choose the
second approach. First, we must seek a model which allows us to pose
quantitative questions. Optimally, we would want to consider a
Hamiltonian which is the sum of one--body and two--body operators. The
one--body terms would describe a nearly degenerate set of
single--particle or shell--model orbitals, and the two--body terms
would contain the random two--body matrix elements. This kind of model
does not allow for an analytical treatment, however, and it is
difficult to see how numerical calculations on the required scale
could be performed. Our work is, instead, based on a Fock--space model
commonly used in the statistical theory of nuclear reactions. We
recall that in nucleon--induced nuclear reactions, precompound
reactions are important whenever the equilibration time of the
compound system is comparable to or even larger than the decay time of
the system. This situation occurs typically at bombarding energies of
several 10 MeV. Then, the standard compound--nucleus model (which uses
the GOE) fails to apply and is replaced by a hierarchical statistical
model: The incident nucleon creates a sequence of $m$--particle
$(m-1)$--hole states where the integer $m$ increases by one unit in
each two--body collision of the projectile with one of the target
nucleons. The density of $m$--particle $(m-1)$--hole states at fixed
excitation energy grows very strongly with increasing $m$, see
Eq.~(\ref{eq7}) and the text following it. Therefore, a dynamical
description of the process is out of the question and statistical
concepts are used instead. We define classes of states.  The states in
class $m$ are composed of the $m$--particle $(m-1)$--hole states.
Within each class, the two--body interaction mixes the $m$--particle
$(m-1)$--hole states. We assume that the resulting matrix problem in
each class $m$ can be replaced by a GOE.  The GOE's pertaining to
different classes (different $m$ values) are uncorrelated. For each
$m$, the one parameter of the GOE is fixed by the density of states in
that class. Classes differing in $m$ by one unit are connected by the
elements of the two--body interaction which are taken to be
independent Gaussian distributed random variables.  This last
assumption corresponds to Feshbach's ``chaining hypothesis''.  It
embodies the two--body character of the interaction and defines a
hierarchical model. Within this model, we ask whether the eigenstates
of the full Hamiltonian are localized. We now turn to the details of
this model. We closely follow the work of Ref.~\cite{nvwy}.

We introduce a basis $|m \mu\rangle$ of states where $m$ is the class
index introduced above, and where $\mu$ with $\mu = 1,\ldots,N_m$ is a
running index. The Hamiltonian reads
\begin{equation}
\label{eq1}
H = \sum_{mn\mu\nu} |m\mu\rangle H_{m\mu,n\nu} \langle n\nu|
\end{equation}
and
\begin{equation}
\label{eq2}
H_{m\mu,n\nu} = \delta_{mn}\delta_{\mu\nu}\ h_m + \delta
H_{m\mu,n\nu} 
\end{equation}
is real and symmetric under exchange of ($m \mu$) and ($n\nu$). We
assume that all matrix elements are Gaussian distributed random
variables with the following first and second moments,
\begin{eqnarray}
\overline{H_{m\mu,n\nu}} & = & \delta_{mn}\delta_{\mu\nu}\ h_m \ , \\
\overline{\delta H_{m\mu,n\nu} \delta H_{m'\mu',n'\nu'}} & = & M_{mn}\ 
(\delta_{mm'}\delta_{nn'}\delta_{\mu\mu'}\delta_{\nu\nu'} +
\delta_{mn'}\delta_{nm'}\delta_{\mu\nu'}\delta_{\nu\mu'})
\label{M-matrix} \ .
\end{eqnarray}
The bar indicates ensemble averaging. For fixed $m$, the matrix
$H_{m\mu,m\nu}$ belongs to the GOE. The matrices describing different
classes $m$ are seen to be uncorrelated. In order to keep the GOE
spectrum finite in the limit $N_m \rightarrow \infty$, the diagonal
terms $M_{mm}$ must scale as $N_m^{-1}$. We consider the weak coupling
case~\cite{nvwy} where the non--diagonal elements $M_{mn}$ ($m \neq
n$) are suppressed by an additional factor $N_n^{-1}$,
\begin{equation}
M_{mn} = \frac{\lambda_m^2}{N_m} \delta_{mn} +
\frac{\lambda_{mn}^2}{N_mN_n}(1-\delta_{mn}) \,.
\label{M_matrix}
\end{equation}
Here $\lambda_m$ and $\lambda_{mn}$ are the strength of the matrix
elements in class $m$ and the coupling strength of states in
different classes, respectively. In keeping with the remarks made
above, we assume that $\lambda_{m n}$ vanishes for $|m - n| \geq
2$. The spectrum of the $N_m$ states in class $m$ has the shape of a
semicircle with center $h_m$ and radius $2 \lambda_m$. We proceed by
assuming that all semicircles have identical centers, $h_m = 0$ for
all $m$, and identical radii, $\lambda_m =\lambda $ for all $m$. The
remaining parameters $N_m$ allow us to account for the fact that the
local level densities $\rho_m(E)$ at energy $E$ differ in the
classes. We work at the centers of the semicircles and put $E = 0$
without loss of generality. Later we need the inverse $g$ of $M$ given
by
\begin{equation}
\label{eq3}
g_{mn} = \frac{N_m}{\lambda^2}\delta_{mn} -
\frac{\lambda_{mn}^2}{\lambda^4}(1-\delta_{mn}) +
O\left(\frac{1}{\sqrt{N_m N_n}}\right)\,.
\end{equation}

With $D(E^\pm) = E - H \pm i\eta$ and the ensemble--averaged
two--point correlation function $\Kscr$ defined by ($m \neq n$)
\begin{equation}
\label{eq4}
\Kscr_{m\mu,n\nu} = \overline{\langle m\mu | D^{-1}(E^+) | n\nu
  \rangle \langle n\nu | D^{-1}(E^-) | m\mu \rangle} \ ,
\end{equation}
the quantity of central interest is the class average of $\Kscr$,
\begin{equation}
\label{eq5}
\Kscr_{mn} = \frac{1}{N_mN_n} \sum_{\mu\nu} \Kscr_{m\mu,n\nu} \,.
\end{equation}
If $\Kscr_{mn}$ vanishes exponentially with increasing distance
$|m-n|$, the eigenstates of $H$ are (exponentially) localized.

The explicit dependence of $\Kscr_{mn}$ on $|m-n|$ can be found with
the help of the supersymmetric non--linear sigma
model~\cite{efetov1,efetov2,vwz} the use of which is by now completely
standard. Moreover, a derivation taylored to the present problem may
be found in Ref.~\cite{nvwy}. Therefore, we restrict ourselves to the
essential steps and emphasize those aspects which are specific to the
present problem. We use the notation of Ref.~\cite{vwz}. The
correlation function is expressed in terms of the generating function
$Z$,
\begin{equation}
\Kscr_{mn} = \left.\frac{\partial^2\ \overline{Z}(J)}{\partial
    J_m^{15}\ \partial J_n^{15}}
\right\arrowvert_{J=0}
\label{corr_funct_deriv_eq}
\end{equation}
where $J_m^{15}$ denotes the (1,5) element (with respect to the
supersymmetry indices) of the auxiliary $J$--field. Similar notation
is used in Eq.~(\ref{eq10}). Here we have averaged the generating
function over the Gaussian distribution of $H_{m\mu,n\nu}$. After
eliminating the quartic dependence on the original integration
variables by means of a Hubbard--Stratonovitch transformation, we find
for $\overline{Z}$ the expression
\begin{eqnarray}
\lefteqn{\overline{Z}(E,J) =}\nonumber\\
& & \int d[\sigma]\ 
\exp\left(-\frac{1}{4}\sum_{mn} \lambda^2 \,
  g_{mn}\,\trg (\sigma_m\sigma_n) - 
\frac{1}{2}\,\trg_{m,\mu,\alpha}\,\ln\,{\mathcal N}(J)\right) 
\label{hstransformed_genfunc} \ .
\end{eqnarray}
The fields $\sigma_m$ are $8\times8$ graded matrices and $d[\sigma] =
\prod_m d[\sigma_m]$. Moreover,
\begin{equation}
{\mathcal N}(J) = E + i \eta + J - \Sigma\,, \ \
\Sigma = \{ \lambda\,\delta_{mn}\,\delta_{\mu\nu}\, 
\sigma^{\alpha\beta}_m\}\,. 
\end{equation}
The integral in Eq.~(\ref{hstransformed_genfunc}) is evaluated by
means of a saddle--point approximation by varying the $\sigma_m$'s. In
view of the smallness of the terms with $m \neq n$ in Eq.~(\ref{eq3}),
we obtain a separate saddle--point equation $\sigma_m (E - \lambda
\sigma_m) = \lambda$ for each class $m$. The solution yields the
semicircle for the level density. Integrating over the massive modes
in the limit where $N_m \rightarrow \infty$ leaves us only the
Goldstone modes $\sigma_m^{\rm G}$. We focus attention on the
correlation function $\Kscr(m) = \Kscr_{1 m}$ which takes the form
\begin{equation}
\Kscr(m) = \frac{1}{2\lambda^2} \int \prod_{k=1}^{M} d[\sigma_k]
(\sigma_1^{\rm G})^{51}(\sigma_m^{\rm G})^{51} \exp\left(
  \sum_{i=1}^{M-1} \frac{\lambda_{i i+1}^2}{4\lambda^2} \trg
  (\sigma_i^{\rm G} \sigma_{i+1}^{\rm G})\right)\,.
\label{corr_function}
\end{equation}
Here $M$ is the total number of classes. In the case of weak coupling
we have~\cite{nvwy}
\begin{equation}
4 \frac{\lambda_{mn}^2}{\lambda^2} = 2 \pi\ \rho_m\
\overline{{V_{mn}}^2}\ 2 \pi\ \rho_n
\end{equation}
where we wrote $M_{mn} = \overline{{V_{mn}}^2}$ and where $\rho_m =
N_m/\pi \lambda$ is the density of states in class $m$ at energy $E =
0$. The form of the expression in the exponent of
Eq.~(\ref{corr_function}) displays the chaining hypothesis (only
states belonging to next--neighboring classes are coupled by
non--vanishing matrix elements).

To proceed, we omit from now on the index G on the sigma fields and
follow Refs.~\cite{zirnbauer,mmgz}. The aim consists in replacing the
summation over classes by an integration (continuum limit). We put
\begin{equation}
\frac{\lambda_{i i+1}^2}{4\lambda^2} = \frac{1}{4\epsilon}\,\alpha_{i,i+1}
\label{prefactor}
\end{equation}
with $\epsilon = 1/\zeta$ and both $\zeta$ and $\alpha_{i,i+1} \gg 1$.
This is justified because the left--hand side of Eq.~(\ref{prefactor})
is much larger than unity, see Eqs.~(\ref{eq9}) and (\ref{eq7}). We
write the correlation function in the form
\begin{equation}
\Kscr(m) = \frac{1}{2\lambda^2}\, \int  d\sigma_1 d\sigma_m\, \sigma^{51}_1\,
W(\sigma_1,\sigma_m;1,m)\, \sigma^{51}_m\, Y(\sigma_m;m)
\label{eq10}
\end{equation}
where
\begin{eqnarray}
W(\sigma_m,\sigma_n;m,n) & = & \int \prod_{k=m+1}^{n-1} d\sigma_k
\exp\left[\,\frac{1}{4\epsilon}\sum_{i=m}^{n-1}\, \alpha_{i,i+1}\,
  \trg\,(\sigma_i\, \sigma_{i+1})\right]\,,\nonumber\\
Y(\sigma_m;m) & = & \int \prod_{k=m+1}^{M} d\sigma_k
\exp\left[\,\frac{1}{4\epsilon}\sum_{i=m}^{M-1}\, \alpha_{i,i+1}\,
  \trg\,(\sigma_i\, \sigma_{i+1})\right] \ .
\end{eqnarray}
Obviously, $W$ obeys the equation
\begin{eqnarray}
\lefteqn{W(\sigma_m,\sigma_{n+1};m,n+1)}\nonumber\\
& &\hspace{20mm} = \int d\sigma_n W(\sigma_m,\sigma_n;m,n)
\exp\left[ 
\,\frac{1}{4\epsilon} \alpha_{n,n+1}\,
  \trg\,(\sigma_n\, \sigma_{n+1})\right]\,.
\label{recursive_eq}
\end{eqnarray}
We use the smallness of $\epsilon$ to take the continuum limit and
replace the summation over classes by an integration over the
continuous variable $t=m/\zeta$. Then,
\begin{eqnarray*}
  \alpha_{m,m+1}           & \rightarrow & \alpha(t)               \\
  W(\sigma_1,\sigma_m;1,m) & \rightarrow & W(\sigma_1,\sigma(t);t) \\
  Y(\sigma_m;m)            & \rightarrow & Y(\sigma(t);t)          \\
\end{eqnarray*}
and Eq.~(\ref{recursive_eq}) becomes an integral equation. This
integral equation is equivalent to a modified heat equation,
\begin{equation}
\alpha(t)\,\partial_t W(\sigma',\sigma;t) = \Delta_\sigma
W(\sigma',\sigma;t)
\label{heat_eq}
\end{equation}
with the initial condition
\begin{equation}
\lim_{t\rightarrow0}\, W(\sigma',\sigma;t)=\delta(\sigma'-\sigma) \,.
\label{W_condition}
\end{equation}
Here $\Delta_\sigma \equiv \partial^2/\partial\sigma^2$
\cite{zirnbauer,mmgz}. 

In previous work~\cite{efetov1,mmgz}, the prefactor in the exponent of
Eq.~(\ref{corr_function}) was a constant (independent of the class
index $i$), and the solution of the heat equation did display an
exponential decay signalling localization. In the present case,
$\alpha(t)$ does depend upon $t$ because the level densities $\rho_m$
depend strongly upon $m$. This fact raises the following questions:
(i) Is localization affected or even destroyed by a monotonic increase
of $\rho_m$ with $m$ and, thus, of $\alpha(t)$ with $t$? (ii) If so,
can we determine the critical exponent $\beta$ in $\alpha(t) \propto
t^{\beta}$ which marks the transition from the localized to the
delocalized regime? (iii) What are the implications of our findings
for localization in nuclei, i.e., for a realistic modelling of the
$t$--dependence of $\alpha(t)$?

To answer these questions, we observe that we retrieve the standard
heat equation by a change of variable,
\begin{equation}
\alpha(t)\, \partial_t = \partial_s \ .
\label{reparameterization_eq}
\end{equation}
In terms of the rescaled function
\begin{equation}
\tilde{W}(\sigma_1,\tilde{\sigma}(s);s) =
\tilde{W}(\sigma_1,\tilde{\sigma}(s(t));s(t))=W(\sigma_1,\sigma(t);t)
\end{equation}
this leads to
\begin{equation}
\partial_s \tilde{W}(\tilde{\sigma}',\tilde{\sigma};s) =
\Delta_{\tilde{\sigma}} 
\tilde{W}(\tilde{\sigma}',\tilde{\sigma};s) \ .
\end{equation}
We choose $s(t)$ such that $s(0)=0$ which ensures that $\tilde{W}$
also obeys Eq.~(\ref{W_condition}). The
transformation~(\ref{reparameterization_eq}) reduces our problem to
the heat equation solved in Ref.~\cite{efetov1}. The rescaled
propagator
\begin{equation}
\tilde{\Kscr}(s) = \frac{1}{2\lambda^2}\, \int
d\sigma_1\,d\tilde{\sigma}(s)\,
\sigma_1^{51}\, 
\tilde{W}(\sigma_1,\tilde{\sigma}(s);s)\,\tilde{\sigma}^{51}(s)\,
\tilde{Y}(\tilde{\sigma(s)};s) 
\label{rescaled_corr_function}
\end{equation}
can be worked out by introducing the Fourier transform $\tilde{K}(k)$
of Eq.~(\ref{rescaled_corr_function}). This yields
\begin{equation}
\tilde{\Kscr}(s) = \frac{\pi^4 \sqrt{\pi}}{16 \zeta\lambda^2}
  \left(\frac{4\zeta}{s}\right)^{3/2} \exp\left(\frac{-s}{4\zeta}\right)
\label{final_rescaled_corr_function}
\end{equation}
which is valid for $s\gg\zeta$. The function $\Kscr(t)$ is obtained by 
solving for $s(t)$ the differential equation
\begin{equation}
\label{eq6}
\frac{d s(t)}{d t} = \frac{1}{\alpha(t)}\,,
\end{equation}
and replacing $s$ by $s(t)$ in
Eq.~(\ref{final_rescaled_corr_function}). For a power--law dependence,
$\alpha(t) \propto t^{\beta}$, Eq.~(\ref{eq5}) yields $s \propto
t^{1-\beta}$. Combining this with
Eq.~(\ref{final_rescaled_corr_function}), we see that $\beta = 1$ is
the critical exponent which marks the transition from localization to
delocalization. The existence of such an exponent is intuitively
clear: A monotonic increase of $\alpha(t)$ with $t$ signals an ever
increasing phase space. The resulting drag on the system
overcompensates, for $\beta > 1$, the tendency of the system to form
localized states.

In order to apply this result to the case of nuclei, we must determine
the dependence of $\alpha(t)$ on $t$. We recall that
\begin{equation}
\alpha(t) \Leftrightarrow  \alpha_{m,m+1}
          =  (\pi^2\, \rho_m\,\overline{{V_{m,m+1}}^2}\,\rho_{m+1}) / \zeta
          =  (\pi^2\, \rho_m\,\Gamma_{m\rightarrow m+1}^\downarrow) /
          \zeta 
\label{eq9}
\end{equation}
where $\Gamma_{m\rightarrow m+1}^\downarrow$ is the spreading width
for transitions from class $m$ to class $m+1$. The width
$\Gamma_{m\rightarrow m+1}^\downarrow$ approximately scales with
$1/m$~\cite{herman}. The density of states in class $m$ at excitation
energy $E$ is~\cite{mirlin}
\begin{equation}
\label{eq7}
\rho_m(E) =
\frac{1}{m!(m+1)!(2m)!\Delta}\left(\frac{E}{\Delta}\right)^{2m}\,
\end{equation}
where $\Delta$ is the mean level spacing of the single--particle
states. For fixed $E$, $\rho_m(E)$ grows strongly with $m$ up to a
maximal value of $m$ and then drops quickly. In the present context,
we are not interested in this cut--off because it might mask the
existence of localization. For this reason, we take the limit of high
excitation energy $E/\Delta \gg m$. Then, we can neglect the factorial
factors in Eq.~(\ref{eq7}) as well as the $1/m$ dependence of
$\Gamma_{m\rightarrow m+1}$ and obtain approximately $\alpha_{m,m+1}
\propto (E/\Delta)^{2m}$. This yields
\begin{equation}
\label{eq8}
s(t) \Leftrightarrow 
     s_m \propto 1 - (E/\Delta)^{-2m}\ .
\end{equation}
We observe that $\alpha_{m,m+1}$ grows with $m$ much more strongly
than linearly. As a consequence, $s_m$ is bounded from above and
approaches asymptotically a finite constant. Thus, the correlation
function $\Kscr(t)$ tends to a finite value as $t \rightarrow \infty$.
We conclude that the interaction spreads the states in class 1 over
the entire Hilbert space.

The results of the model investigated in the present paper negate the
possibility of localization in a many--body Fermi system like the
nucleus. This finding must be contrasted with the results of
Ref.~\cite{agkl} where the chain of $m$--particle $(m-1)$--hole states
was mapped onto a Bethe lattice, and localization was predicted. Such
mapping completely neglects the interaction between states within each
class $m$. In contradistinction, our use of a GOE for the states
within each class may overemphasize the role of the interaction
between states within each class $m$, although we believe the present
model to be closer to reality in nuclei than the model using a Bethe
lattice. As mentioned above, in the numerical work of Leyronas et
al.~\cite{leyronas} evidence for localization was presented. In that
work, all those $m$--particle $(m-1)$--hole states were taken into
account the unperturbed energies of which lie in some small
neighborhood of the energy $E$ considered. It is not clear to us
whether this constraint might not affect the localization properties
of the system.

In summary, we have shown that in the framework of our hierarchical
model localization is destroyed whenever the density of states
increases more strongly than linearly with increasing complexity $m$
of states. In the case of nuclei, the model does not predict
localization.

T.R. thanks C. Mej{\'\i}a--Monasterio for beneficial discussions at
the early stages of this work.


\begin{thebibliography}{99}

\bibitem{guhr}
  T.~Guhr, A.~M\"uller--Groeling, and H.A.~Weidenm\"uller,
  Phys. Rep. 299 (1998) 189.  

\bibitem{gervois}
  A.~Gervois, Phys. Lett. B 26 (1968) 413.

\bibitem{flambaum}
  V.V.~Flambaum, A.A.~Gribakina, G.F.~Gribakin, and M.G.~Kozlov,
  Phys. Rev. A 50 (1994) 267. 

\bibitem{french}
  J.B.~French and S.S.M.~Wong, Phys. Lett. B 35 (1971) 5.

\bibitem{bohigas}
  O.~Bohigas and J.~Flores, Phys. Lett. B 35 (1971) 383.

\bibitem{agkl}
  B.L.~Altshuler, Y.~Gefen, A.~Kamenev, and L.S.~Levitov,
  Phys. Rev. Lett. 78 (1997) 2803. 

\bibitem{leyronas}
  X.~Leyronas, J.~Tworzyd{\l}o, and C.W.J. Beenakker, Phys. Rev. Lett. 82
  (1999) 4894. 

\bibitem{mejia}
  C.~Mej{\'\i}a--Monasterio, J.~Richert, T.~Rupp, and
  H.A.~Weidenm{\"u}ller, Phys. Rev. Lett. 81 (1998) 5189. 

\bibitem{anderson}
  P.W.~Anderson, Phys. Rev. 109 (1958) 1492.

\bibitem{wiersma}
  D.S.~Wiersma, P.~Bartolini, A.~Lagendijk, and R.~Righini, Nature 390 
  (1997) 671. 

\bibitem{logan}
  D.E.~Logan and P.G.~Wolynes, J. Chem. Phys. 93 (1990) 4994.

\bibitem{nvwy}
  H.~Nishioka, J.J.M.~Verbaarschot, H.A.~Weidenm\"uller, and
  S.~Yoshida, Ann. Phys. (N.Y.) 172 (1986) 67.

\bibitem{efetov1}
  K.B.~Efetov, Adv. Phys. 32 (1983) 53.

\bibitem{efetov2}
  K.B.~Efetov, Supersymmetry in disorder and chaos (Cambridge
  University Press, Cambridge, 1997). 

\bibitem{vwz}
  J.J.M.~Verbaarschot, H.A.~Weidenm\"uller, and M.R.~Zirnbauer,
  Phys. Rep. 129 (1985) 367. 

\bibitem{zirnbauer}
  M.R.~Zirnbauer, Phys. Rev. Lett. 69 (1992) 1584. 
  
\bibitem{mmgz} 
  A.D.~Mirlin, A.~M\"uller--Groeling, and M.R.~Zirnbauer,
  Ann. Phys. (N.Y.)  236 (1994) 325. 

\bibitem{herman}
  M.~Herman, G.~Reffo, and H.A.~Weidenm\"uller, Nucl. Phys. A 536
  (1992) 124.

\bibitem{mirlin} 
  A.D.~Mirlin and Y.V.~Fyodorov, Phys. Rev. B 56 (1997) 13393.

\end{thebibliography}
\end{document}